\documentclass[aps,prfluids,reprint,groupedaddress]{revtex4-2}
\usepackage{CJK}

\usepackage{natbib}

\usepackage{graphicx}
\usepackage{tabularx}

\begin{document}
\begin{CJK*}{UTF8}{} 
\CJKfamily{gbsn}
\preprint{}

\title{Rayleigh-Taylor Unstable Flames: the Effect of Two-Mode Coupling}


\author{Mingxuan Liu (刘明轩)}
\affiliation{Emory University, Atlanta, Georgia 30322, USA}
\email[]{mgxnliu@umich.edu}
\author{Elizabeth P. Hicks}
\affiliation{Epsilon Delta Labs, Evanston, Illinois 60201, USA}
\email[]{ephicks@fastmail.net}

\collaboration{Accepted by Physical Review Fluids on November 5, 2024}


\begin{abstract}
In the classical Rayleigh-Taylor (RT) instability, initial conditions are forgotten and the growth of the mixing layer becomes self-similar when short wavelength modes couple to generate longer wavelength modes. In this paper, we explore how adding a reaction at the unstable interface affects this inverse cascade in wavenumber (``inverse k-cascade''). We simulate a 2D, Boussinesq, premixed model flame perturbed by a large amplitude primary mode ($k_1$) and a smaller amplitude secondary mode ($k_2$). Early on, the modes are uncoupled and the flame propagates as a metastable traveling wave. Once the secondary mode has grown large enough, the modes couple. The traveling wave is destabilized and the flame front bubbles rapidly grow. This inverse k-cascade, driven by two-mode coupling, ultimately generates a long wavelength mode with wavenumber GCD$(k_1,k_2)$, where GCD is the greatest common divisor. We identify five distinct flame growth solution types, and show that the flame may stall, develop coherent pulsations, or even become a metastable traveling wave again depending on GCD$(k_1,k_2)$. Finally, we compare our results with two-mode coupling in ablative and classical RT and show that all three systems may follow the same mode coupling dynamics.
\end{abstract}


\maketitle
\end{CJK*}

\onecolumngrid

\section{Introduction} \label{introduction}

The classical Rayleigh-Taylor (RT) instability \citep{rayleigh1883,taylor1950} occurs when a heavy fluid is accelerated into a light fluid. This acceleration may be due to gravity or even the centrifugal force. The RT instability is very well studied \citep{sharp1984, bychkov2007, andrews2010, anisimov2013, livescu2013, boffetta2017, zhou2017a, zhou2017,zhou2019, banerjee2020, livescu2020, schilling2020, zhou2021} and appears in many places from the Crab Nebula \citep{hester1996,porth2014} to Earth's ionosphere \citep{shinagawa2018}.  In some important applications, there is an additional twist: a reaction at the interface between the heavy and light fluids. For example, the speed of thermonuclear flames in Type Ia supernovae is increased by the RT instability \citep{khokhlov1994, khokhlov1995, V03, V05, bell2004b, zingale2005a, zhang2007, chertkov2009, biferale2011, hicks2013, hicks2015, hicks2019, ciaraldi-schoolmann2009, hristov2018}. Here on Earth, engineers seek to improve the efficiency of aviation gas turbine engines by using the RT instability to speed up fuel consumption \citep{lewis1971, lewis1973, lewis1975,lapsa2009, briones2015, erdmann2019, sykes2019, sykes2021, erdmann2023}. 

One characteristic feature of classical RT is that initial conditions are forgotten as the growth of the mixing layer becomes self-similar.  Memory is lost when long wavelength modes are generated by the coupling of short wavelength modes (``bubble merger'') instead of growing from their own initial conditions (``bubble competition'') \citep{birkhoff1955, sharp1984, haan1989, haan1991, alon1993, alon1994, alon1995, shvarts1995, dunning1995, ofer1996, oron2001, cheng2002, dimonte2004a, dimonte2005}. Many generations of bubble merger generate an inverse k-cascade in wavenumber space as bubbles grow from small to large. In this paper, we explore how the addition of a reaction affects the bubble merger process. 

The development of long wavelength modes in reactive RT has never been explicitly studied.   Previous work has focused on the long-term development of the instability \citep{chertkov2009, ley2024} and on measuring properties like the flame speed and flame width \citep{khokhlov1994, khokhlov1995, V03, V05, bell2004b, zingale2005a, zhang2007, chertkov2009, biferale2011, hicks2013, hicks2015, hicks2019}. Generally, we might expect bubble merger in reactive RT to resemble bubble merger in either classical RT or ablative RT. In both cases, the dominant modal interaction is two-mode coupling; we expect this also to be the case for reactive RT. In classical RT, the Haan/Ofer model predicts that two short wavelength modes will couple to generate a new long wavelength mode with wavenumber $k_{new}=|k_1 - k_2|$ \citep{haan1991, shvarts1995, ofer1996}. In ablative RT, Xin's study \citep{xin2019} of two-mode coupling produced a surprising result: a new mode with $\lambda_{new}=LCM(\lambda_1,\lambda_2)$, where $LCM$ is the least common multiple. Do two-mode interactions in reactive RT more closely resemble those of classical RT or ablative RT?

In this paper, we investigate two-mode coupling for reactive RT using numerical simulations. Given that this topic is unexplored, we begin by answering some basic questions: Can two-mode coupling produce long wavelength modes? Which long wavelength modes are generated? What happens to the flame physically during the inverse k-cascade? Ultimately, we will show that reactive RT follows a similar coupling dynamic to the ablative RT, but that this seemingly strange mode coupling dynamic can emerge from the Haan/Ofer model.

\section{Numerical Methodology} \label{numerics}

In this study, we take a parameter study approach. We reduce computational expense and increase our exploration space by making two major simplifications: the Boussinesq approximation and a model reaction. 

The Boussinesq approximation assumes subsonic flows with small density variations \citep{spiegel1960}. For flames, it requires the Atwood number of the system, $At = (\rho_0 - \rho_1)/(\rho_0 + \rho_1)$, to be much less than $1$. Here, $\rho_0$ is the density of the fuel and $\rho_1$ is the density of the ash. By using the Boussinesq approximation, we remove other instabilities (e.g. the Landau-Darrieus instability) from the problem and focus on the effects of the Rayleigh-Taylor instability. 

The second important simplification is the use of a model reaction term, $R(T)$, instead of a realistic chemical reaction network. The reaction progress variable $T$ (temperature) tracks the transformation of fluid from unburnt fuel ($T=0$) to burnt ashes ($T=1$) and represents the amount of energy released from the reaction to the flow \citep{vladimirova2006}. We adopted $R(T)=2 \gamma T^2(1-T)$, a model reaction used in our previous studies of Rayleigh-Taylor unstable flames \citep{hicks2013, hicks2014, hicks2015, hicks2019}. This model reaction has a laminar solution with characteristic flame width $\delta$ and laminar flame speed $s_o$ \citep{constantin2003}. Using the thermal diffusivity $\kappa$ and the laminar reaction rate $\gamma$, we can construct the laminar flame width $\delta = \sqrt{\kappa / \gamma}$ and the laminar flame speed $s_o = \sqrt{\gamma \kappa}$. The fluid equations, non-dimensionalized by $\delta$ and $s_o$, are

\begin{equation}
\frac{D\textbf{u}}{Dt} = -\left(\frac{1}{\rho_0}\right) \nabla p + GT + Pr \nabla^2 \textbf{u} 
\end{equation}
\begin{equation}
\frac{DT}{Dt} = \nabla^2 T + 2T^2(1-T),
\end{equation}
and $\nabla \cdot  \textbf{u} = 0$. $G = g (\Delta \rho / \rho_0) (\delta / s^2_0)$ is the non-dimensionalized gravity, $Pr = \nu / \kappa$ is the Prandtl number, and $\Delta \rho$ is the density jump across the flame front. We set $G=4$ and $Pr=1$ for all simulations in order to be within the range of values considered by previous work \citep{V03,V05}. 

We chose 2D simulations because they are less computationally expensive than 3D, allowing us to probe more parameter space. Does this expedient choice make our simulations too physically unrealistic to be useful? It depends on what we study. Here, we investigate mode coupling and phenomenological solution types. We expect our 2D results to have qualitative counterparts in 3D because, like in classical RT, the inverse cascade in k-space is driven by the physical merger of bubbles, not by turbulence. The physical merger of bubbles is qualitatively similar in 2D and 3D. Quantitative differences are due to the differences between 2D and 3D bubbles and include different drag and mass coefficients, a higher asymptotic velocity for 3D bubbles, and tighter packing and more elongation of 3D bubbles \citep{oron2001}. General differences between 2D and 3D classical RT are reviewed by \citet{zhou2017}.

To solve the fluid equations, we ran direct numerical simulations using Nek5000, an open-source spectral element CFD code \citep{nek5000}. Our simulations have a physical size of $2048$ x $9216$ non-dimensionalized units, spanned by $512$ x $2304$ elements with a spectral order of $N=9$ (see Appendix \ref{resolution} for resolution tests). Periodic boundary conditions were imposed in the $x$-direction. Simulating a tall box ensured that the flame stayed well away from the top and bottom boundaries during its evolution. The velocity field was initialized at $0$ and was held at $0$ on both the top and bottom boundaries.  The temperature was constrained to be $0$ (fuel) at the top boundary and $1$ (ash) at the bottom boundary. The flame front starts in the middle of the box ($y_0=4608$) and propagates upwards against gravity.

We consider the coupling between a large amplitude primary mode and a small amplitude secondary mode. Both modes are sinusoidal. Initially, the amplitude of the flame front's deviation from flat is
\begin{equation}
h(x) = A_1 \sin\left(\frac{2\pi k_1 x}{x_{\rm max}} \right) + A_2 \sin\left(\frac{2\pi k_2 x}{x_{\rm max}} \right), 
\end{equation}
where $A_1$ ($A_2$) and $k_1$ ($k_2$) are the amplitude and wavenumber of the primary (secondary) perturbation, and $x_{\rm max}=2048$ is the box width. $A_1=1$ and $k_1=128$ for all simulations. $A_2=0.001$ for most simulations and $k_2$ ranges from 1 to 224. In addition, we explicitly consider the effect of numerical multimode noise in our simulations. For a discussion see Appendix \ref{resolution}. The initial temperature profile is $T(x,y) = 0.5\{1 - \tanh[(y - y_0 + h(x))/2] \}$. For movies of the temperature field for all simulations, see the Supplemental Materials \citep{liu-supplement}. 

\section{Results} \label{results}
To investigate how long wavelength modes are generated by two-mode coupling in reactive RT, we define two stages of flame evolution, Early and Late, depending on whether or not the primary and secondary modes are coupled. During the Early Stage, the primary perturbation grows and then stabilizes into a traveling wave. The Late Stage begins when the travelling wave is disrupted by the growing secondary perturbation (or by numerical multimode noise) and mode coupling drives an inverse k-cascade towards long wavelengths. We explore the flame's physical behavior during the inverse k-cascade and show that the cascade ultimately generates a dominant mode with wavenumber GCD$(k_1,k_2)$. Finally, we compare two-mode coupling in reactive, ablative, and classical RT.

\begin{figure*}[hbt!]
  \includegraphics[width=0.49\linewidth]{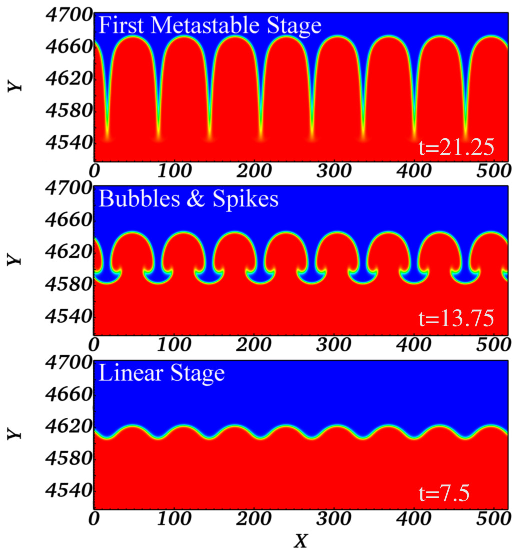}
\caption{The Early Stage: the initial perturbation grows exponentially during the linear stage (bottom panel); bubbles and spikes develop (middle panel); finally, the flame stabilizes as a traveling wave that we call the ``First Metastable Stage'' (top panel). This is a single-mode simulation with $k_1=32$; see the Supplemental Material Movie for run 355 \citep{liu-supplement}.}
\label{fig1}
\end{figure*}
\vspace{-0.75cm}
\subsection{Early Stage Flame Evolution} \label{earlystage}
We begin by exploring how the primary and secondary modes grow independently during the earliest stages of flame evolution.

The flame begins as a slightly perturbed sine wave dominated by the primary perturbation.  The primary perturbation grows exponentially (see Figure \ref{fig1}, bottom panel) during the linear growth stage \citep{zeldovich1985, V03}, but this growth slows as the flame transitions into the nonlinear regime.  The less dense ash rises as bubbles and the more dense fuel sinks as spikes.   The spikes have a more complex structure when the RT instability is stronger; for example, they may resemble mushrooms (see Figure \ref{fig1}, middle panel).  Finally, the fine structure on the spikes burns out and the flame becomes a regular series of rising bubbles separated by sharp cusps (see Figure \ref{fig1}, top panel) \citep{V03, V05}. We call this flame configuration the ``First Metastable Stage.''

During the First Metastable Stage, the flame propagates upwards with a constant speed, shape and phase, that is, as a traveling wave \citep{bayliss2001,V03,V05}. Most properties of the First Metastable Stage depend on the dominant primary perturbation. For example, the number of RT bubbles in the box is equal to the primary wavenumber $k_1$. The flame speed and cusp size depend on both $k_1$ and $G$ \citep{khokhlov1995}. However, the lifetime of the First Metastable Stage depends on how long it takes the secondary perturbation (or numerical multimode noise; see Appendix \ref{resolution}) to grow and disrupt it. This depends on the properties of the secondary perturbation.

\begin{figure*}[hbt!]
  \includegraphics[width=0.97\linewidth]{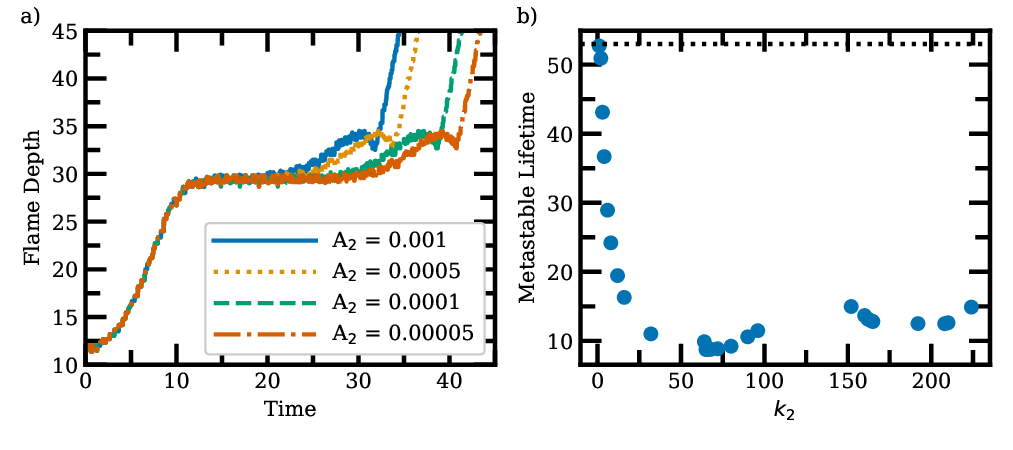}
\caption{(a) Secondary Perturbation Amplitude Experiment: the lifetime of the First Metastable Stage increases as $A_2$ decreases. On the plot, the First Metastable Stage is the plateau where the flame depth remains constant. The flame depth is measured with the point method using $T=[0.005, 0.995]$ (see Appendix \ref{depth}). (b) Secondary Perturbation Wavenumber Experiment: the lifetime of the First Metastable Stage decreases with $k_2$ and then stabilizes. The noise lifetime for $k_1=128$ (see Appendix \ref{resolution}) is represented by the dotted line.}
\label{fig2}
\end{figure*}

For instance, the lifetime of the First Metastable Stage depends on the amplitude of the secondary perturbation. Fixing $k_1=128$, $k_2=64$, we ran simulations with $A_2=0.001,0.0005,0.0001,0.00005$. Figure \ref{fig2}a shows how the distance between the top of the bubbles and the bottom of the spikes (the ``flame depth'', see Appendix \ref{depth}) changes with time. After growing exponentially, the flame depth remains constant during the First Metastable Stage. Eventually, the secondary perturbation grows large enough to couple with the primary perturbation, and the flame depth increases again. Figure \ref{fig2}a shows that smaller secondary perturbations take longer to grow, increasing the lifetime of the First Metastable Stage. 

The lifetime of the First Metastable Stage also depends on the wavenumber of the secondary perturbation.  Setting $k_1=128$, $A_1=1$, $A_2=0.001$ we chose $k_2$ from the range 1 to 224 and then measured the lifetime of the First Metastable Stage. Figure \ref{fig2}b shows that the lifetime of the First Metastable Stage is very long ($t_{\rm life}>50$) when $k_2$ is small, but rapidly drops as $k_2$ increases, and then varies within a band $8.7 \leq t_{\rm life} \leq 15$.  The rapid drop in lifetime is due to the secondary perturbation growing faster as $k_2$ increases. This is consistent with the classical RT linear growth rate scaling, which is proportional to $\sqrt{k}$. Lifetimes stabilize at higher $k_2$ because the growth rate of the secondary perturbation levels off.  This happens because the reaction, thermal diffusion, and viscosity all destroy the smaller scale structures of the higher $k_2$ perturbations more effectively, offsetting the $\sqrt{k}$ growth rate scaling. So, the lifetime of the First Metastable Stage does depend on $k_2$, but this dependence is strongest when $k_2$ is small.

In the Early Stage, the primary saturates as a traveling wave that persists until the independently growing secondary perturbation (or multimode noise) grows large enough to couple with it. In this section, we showed that the secondary's growth time depends on its initial amplitude and wavenumber.
\vspace{-0.2cm}
\subsection{Late Stage Flame Evolution} \label{latestage}

In the Late Stage, the primary and secondary modes couple to generate intermediate and long wavelength modes. In this section, we will show that the final mode generated by this inverse k-cascade is GCD$(k_1,k_2)$, where GCD is the greatest common divisor. We also will explore the variety of solutions that emerge during the inverse k-cascade, from disordered chaotic burning to structured symmetrical evolution, and show that they depend on GCD$(k_1,k_2)$. 

The Late Stage begins when the primary perturbation's First Metastable Stage is destabilized by another perturbation, either the secondary perturbation or numerical multimode noise. We've shown in Section \ref{earlystage} that the lifetime of the First Metastable Stage depends on $k_2$ and that long wavelength secondary perturbations grow very slowly. For $k_2=1,2$ the secondary perturbation grows too slowly to disrupt the First Metastable Stage before the numerical multimode noise kicks in.  In this \textbf{Chaotic Burning} solution (see Figure \ref{chaoticburning} in Appendix \ref{solutions} for $k_2=1$), the First Metastable Stage (Figure \ref{chaoticburning}, panel 1) is disrupted by small scale asymmetrical bubble growth (panel 2). Here and there, a few bubbles along the flame front randomly grow, absorbing their neighbors (panels 2-4). This leads to rapid merging and growth of the bubbles (panels 5-8). The vertical growth of the bubble mixing layer (measured in several ways, see Appendix \ref{depth}) is shown in the inset ``bubble depth'' plot.

Moving upwards in wavenumber, a new solution type emerges at $k_2=3$. \textbf{Nearly Symmetric Merging} solutions feature a nearly symmetric breakup of the First Metastable Stage, but small asymmetries are magnified before wavenumber GCD$(k_1, k_2)$ structures are reached, leading to asymmetric bubbles.  We see this solution type for $k_2=3,4$ and for high $k_2$ wavenumbers with GCD$=1$, specifically $k_2=65,165$. Figure \ref{nearlysymmetricmerging} shows $k_2=65$.

Continuing to higher wavenumbers, when $k_2 \ge 8$, the secondary perturbation outcompetes the multimode noise and delays the emergence of asymmetry. In these simulations, the solution type is determined entirely by the wavenumber of the mode ultimately produced by two-mode coupling: GCD$(k_1, k_2)$. This mode divides the domain into GCD$(k_1, k_2)$ equal subregions and identical evolution takes place within each one. Bubbles are distributed evenly into the subregions and merge until only one bubble remains per subregion. Thus, the final wavenumber is GCD$(k_1, k_2)$. The details of the merger process depend on $k_2$ and GCD$(k_1, k_2)$. If $k_2=$ GCD$(k_1, k_2)$, then $k_2$ grows and outcompetes $k_1$ in a bubble-competition-like process. Smaller primary bubbles merge on the surface of larger, faster-growing secondary bubbles. A similar bubble-competition-like process takes place for certain other values of $k_2$ if a long wavelength mode is quickly generated. Although these details of the merger process depend on $k_2$, the solution type only depends on the GCD.

At low GCD (GCD=$2,4$) and $k_2 \ge 8$, we find the \textbf{First Metastable Merging} solution.  In this solution type, bubbles continuously merge while keeping remarkable symmetry, despite their vast structures and high flame speeds. For example, Figure \ref{firstmetastablemerging} ($k_2=210$, GCD=$2$) shows smaller bubbles (panel 2) merging into larger bubbles (panels 3-7) and eventually forming two massive identical structures (panel 8). At the end of the merging process, the GCD=$2$ simulations ($k_2=66, 90, 162, 210$) form two giant identical structures, while the  GCD=$4$ simulations ($k_2=12, 68, 164$) form four. At very late times, multimode noise breaks the symmetry and growth continues.

At medium GCD (GCD=$8,16,32$), the \textbf{First Metastable Pulsating} solution appears. Bubbles merge until only a GCD number of bubbles remain. These bubbles pulsate horizontally without growing vertically as the flame burns upwards. Pulsations are driven by the misalignment between the primary First Metastable Stage and secondary perturbation growing beneath it. When the flame bubble or cusp does not perfectly sit atop the secondary perturbation's crest, horizontal momentum is introduced into the flame front. This causes the bubbles to merge and the flame wakes to pulsate. Figure \ref{firstmetastablepulsating} shows this solution for $k_2=224$ (GCD=$32$). After the First Metastable Stage (panel 1) is broken by the secondary perturbation, bubbles merge (panel 2) until $32$ remain (panel 3). The bubbles pulsate back and forth horizontally, alternately forming and burning out mushroom-like cusps to the right and then to the left (panel 4). Finally, multimode noise disrupts the pulsating solution (panels 5-6) and merger continues to larger horizontal scales (panel 7). The GCD=$32$ simulation pulsations ($k_2=32, 96, 160, 224$) look simple because the $k=32$ structures aren't large. The final structures are larger and have more complex-looking pulsations for GCD=$16$ ($k_2=16, 80, 208$) and GCD=$8$ ($k_2=8, 72, 152$).

The \textbf{Second Metastable Merging} solution requires the highest GCD value (GCD=$64$), half of the primary wavenumber. This solution is special, because the primary and secondary modes interact without introducing horizontal momentum. Figure \ref{secondmetastablemerging} shows this process for $k_2=64$. Exactly two primary metastable bubbles fit within each secondary perturbation wave, so the primary bubbles will alternate between having their upward velocity reinforced and suppressed (panel 2). The reinforced bubbles engulf the suppressed bubbles (panels 3-4) and a Second Metastable Stage forms with $k=64$. The Second Metastable Stage lasts for a considerable amount of time (panels 5-7), gently pulsating due to a shear instability that develops behind the flame front \citep{hicks2014}. Finally, multimode noise triggers chaotic burning (panels 8-9). The $k_2=192$ simulation develops similarly, with two primary metastable bubbles per three secondary wavelengths. Again, alternate primary bubbles are reinforced and suppressed by the secondary perturbation.

\begin{figure*}[hbt!]
\centering
\includegraphics[width=0.81\textwidth]{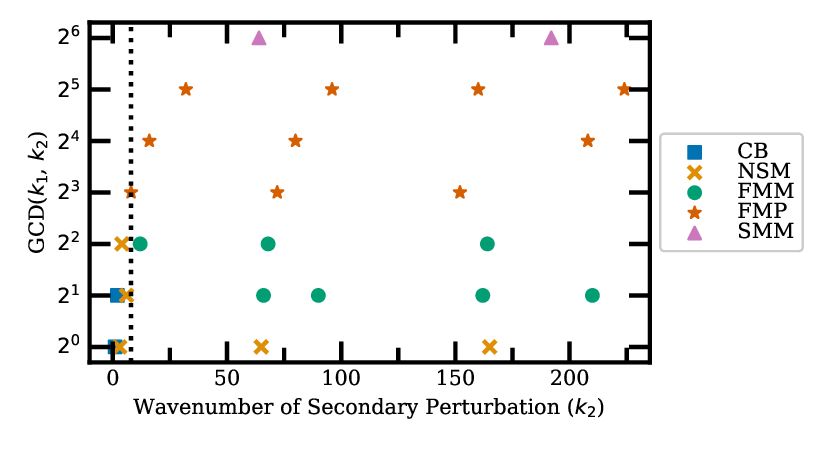}
\caption{Solution Type Phase Diagram. The solution types are Chaotic Burning (CB), Nearly Symmetric Merging (NSM), First Metastable Merging (FMM), First Metastable Pulsating (FMP), and Second Metastable Merging (SMM). Note that the solution type only depends on GCD$(k_1,k_2)$ when $k_2 \ge 8$ (i.e. on and to the right of the dotted line).}
\label{phasediagram}
\end{figure*}

In this subsection, we showed how different Late Stage inverse k-cascade scenarios arise from the coupling between the primary and secondary modes. Figure \ref{phasediagram} summarizes the distribution of the five solutions that we identified in GCD$(k_1,k_2)$ versus $k_2$ phase space. Solutions range from totally disordered to completely ordered. As long as the secondary mode grows quickly enough, the solution type is determined entirely by two-mode coupling, which generates a final mode with wavenumber GCD$(k_1,k_2)$. Visually, this mode appears as a GCD number of identical bubble structures. Whether this mode appears quickly (bubble-competition-like) or results from many generations of bubble merger depends on $k_2$. Ultimately, multimode noise emerges and breaks the symmetry of the GCD subregions, causing continued bubble growth.

\subsection{Two-Mode Coupling: Comparing Reactive, Ablative, and Classical RT} \label{coupling}
In the previous section, we showed that two-mode coupling between the primary and secondary ultimately generates a mode with wavenumber GCD$(k_1,k_2)$. Here, we compare this result to two-mode coupling in ablative and classical RT.

Ablative RT is a close relative of reactive RT. Like reaction, ablation stabilizes linear growth at high $k$ \citep{takabe1991, betti2006, casner2015, martinez2015, yan2016, zhang2018, zhang2018a, xin2019, remington2019}. Ablative RT is important in inertial confinement fusion (ICF) because it inhibits ignition by mixing the cold, dense pellet shell into the fusion fuel. To further understand how initial perturbations affect this mixing, \citet{xin2019} studied two-mode coupling of the ablative RT using 2D direct numerical simulations. They found that two short wavelength modes couple to generate a long wavelength mode that dominates the flow. This new mode has a wavelength LCM$(\lambda_1, \lambda_2)$ which is equivalent to wavenumber GCD$(k_1,k_2)$, our reactive RT result.

Seemingly in contrast, many other systems are driven by a two-mode coupling mechanism that generates the modes $|k_1-k_2|$ and $k_1+k_2$. For example, \citet{newman2000} uses Burgers equation to illustrate how the generic nonlinearity $u u_x$ produces these modes, leading to an inverse k-cascade. In classical RT, Haan's model \citep{haan1991} with Ofer's nonlinear closures \citep{shvarts1995, ofer1996} (which extend the model past the weakly nonlinear state) also generates the $|k_1-k_2|$ and $k_1+k_2$ modes. The $k_1+k_2$ mode is a higher harmonic, while the $|k_1-k_2|$ mode is the start of the inverse k-cascade. As new long wavelength modes are generated, they couple to produce even longer wavelength modes. But what is the final mode generated by this process? As long as noise is sufficiently small and at least one parent mode is able to couple with each new child mode, the result is GCD$(k_1,k_2)$. This is because GCD$(|k_1-k_2|,k_2)=$ GCD$(k_1,|k_2-k_1|)=$ GCD$(k_1,k_2)$, so repeatedly subtracting pairs of modes will eventually result in GCD$(k_1,k_2)$. In essence, this mode coupling mechanism carries out the simplest version of Euclid's algorithm, which finds the GCD of two numbers by repeated subtraction.

Do ablative, reactive, and classical RT all have the same two-mode coupling dynamics? Our results show that this is a possibility -- Xin's surprising LCM$(\lambda_1, \lambda_2)$ final mode and our GCD$(k_1,k_2)$ final mode are consistent with a model that, like the Haan/Ofer model, repeatedly generates new modes with $|k_1-k_2|$. However, the Haan/Ofer model itself does not apply to the reactive case since Ofer's mode saturation closures are inconsistent with our results. In particular, our results in Sections \ref{earlystage} and \ref{latestage} show that saturated reactive modes do not continue to grow, but are able to participate in the inverse k-cascade after saturation. Ultimately, it may be the case that ablative, reactive and classical RT all generate new long wavelength modes by the two-mode coupling rule $|k_1-k_2|$, but that the nonlinear mode saturation closure is different in each case.

\section{Conclusions} \label{conclusions}

In this paper, we investigated how long wavelength modes develop in reactive RT. We focused on two-mode coupling, the dominant modal interaction in both classical and ablative RT. Our main goals were to determine which long wavelength modes are generated, if any, and what happens to the flame physically during the inverse k-cascade. We simulated a 2D Boussinesq model flame with two perturbations: a large amplitude primary and a small secondary. We also considered the role of numerical multimode noise. We divide the flame's evolution into two stages depending on whether or not the modes couple. In the Early Stage, the modes grow independently. The lifetime of the primary's First Metastable Stage depends on the secondary perturbation's initial amplitude and wavenumber. In the Late Stage, the primary and secondary modes couple to ultimately generate a long wavelength mode with wavenumber GCD$(k_1, k_2)$. The flame's behavior during this inverse k-cascade depends on GCD$(k_1, k_2)$. We identified five solution types with various flame dynamics: the flame may be chaotic (if its evolution is dominated by numerical noise), may grow with a slight asymmetry, may symmetrically grow, may coherently pulsate, or may even become metastable again. Of these solutions, the pulsating and second metastable solutions demonstrate how adding a reaction can temporarily stabilize the growth of the mixing layer, an effect not found in classical RT. Finally, we showed that the reactive RT follows a similar coupling dynamic to ablative RT, since both produce modes with GCD$(k_1, k_2)$. We also showed that ablative, reactive, and classical RT may all have the same two-mode coupling dynamics: an inverse k-cascade driven by the generation of new modes with $|k_1-k_2|$. However, the nonlinear mode saturation closure must be different for reactive and classical RT. Ultimately, this work is the first step towards understanding how a reaction modifies the RT bubble merger process and how the mixing layer in reactive RT ultimately becomes self-similar.

\medskip
\noindent\textbf{Data Availability Statement}. The supporting data and code for this article are openly available on Zenodo \citep{liu2024-code}.


\begin{acknowledgments}
E. Hicks thanks R. Rosner for originally introducing her to Rayleigh-Taylor unstable flames and N. Vladimirova and A. Obabko for introducing her to the Nek5000 code and for providing the original RT unstable flames setup and scripts. She also thanks R. Rosner and N. Vladimirova for interesting discussions that have influenced her thinking over the years and T. Erdmann and J. Sykes for a fascinating discussion on the applications of RT unstable flames to aviation. We are very grateful to P. Fischer, A. Obabko, and the rest of the Nek5000 team for making Nek5000 available and for giving us advice on using it. Thank you to S. Tarzia for proofreading and editing suggestions. Thank you to the anonymous referees for their thoughtful comments and advice.

This work used the Extreme Science and Engineering Discovery Environment (XSEDE) \citep{xsede}, which is supported by National Science Foundation grant number ACI-1548562. M. Liu and E. Hicks thank the XSEDE EMPOWER program, supported by National Science Foundation grant number ACI-1548562. This work used resources of the Advanced Cyberinfrastructure Coordination Ecosystem: Services \& Support (ACCESS) program \citep{boerner2023}, which is supported by National Science Foundation grants \#2138259, \#2138286, \#2138307, \#2137603, and \#2138296.  This work used the XSEDE and ACCESS resources Stampede2 and Ranch at the Texas Advanced Computing Center (TACC) through allocation SEE220001. The authors acknowledge the Texas Advanced Computing Center (TACC) at The University of Texas at Austin for providing HPC and visualization resources that have contributed to the research results reported within this paper. URL: http://www.tacc.utexas.edu. This work used the XSEDE and ACCESS resource Expanse at the San Diego Supercomputer Center (SDSC) at the University of California San Diego through allocation SEE220001. Simulations were run using Nek5000 \citep{nek5000}. Simulation visualizations were created using VisIt \citep{visit}. Paper plots were created using matplotlib \citep{matplotlib, matplotlib-3.5.3} and seaborn \cite{seaborn}. Additional plots were made using Gnuplot \citep{gnuplot-5.2.4}. We used Poetry \citep{poetry} for Python dependency management. Our Python analysis code used the packages pandas \citep{pandas, pandas-1.4.4}, NumPy \citep{numpy}, SciPy \citep{scipy}, lmfit \citep{lmfit-1.1.0} and pytest \citep{pytest-7.2.0}.
\end{acknowledgments}

\newpage
\appendix
\vspace{-0.25cm}

\section{Resolution Checks} \label{resolution}

In this appendix, we discuss the effect of numerical errors in our simulations and how we controlled these errors in our study design. We begin by describing the types of numerical error. Next, we explain how we used single-mode simulations and resolution studies to choose parameter values for our two-mode study. Finally, we carefully consider the role of numerical multimode noise in our two-mode simulations.

\subsection{Types of Numerical Error}

The effect of numerical errors in our simulations can be broken into two components: a single-mode ``system'' perturbation and a multimode noise perturbation. The system perturbation is generated by an interaction between the primary perturbation and the Nek5000 spectral element mesh.  If the wavelength of the primary perturbation is not evenly divided by the spectral element size, then the pattern of spectral coefficients repeats on the smallest scale for which the flame front pattern is exactly divided by a whole number of spectral elements. This produces a long wavelength system perturbation with wavenumber GCD$(k_1, nelx)$, where $nelx=512$ is the number of elements in the $x$-direction. The second type of numerical perturbation is multimode noise. Both types of perturbation grow with time. In the next two sections, we will examine the effect of these numerical perturbations on both single-mode and two-mode simulations.

\subsection{Single-Mode Simulations and Resolution Checks}

In this section, we show that single-mode simulations behave nicely with improved resolution. The resolution studies guide our choice of an appropriate primary wavenumber and simulation resolution for our two-mode study. We also measure the ``noise lifetime'' for our chosen $k_1$, $A_1$, and resolution.

\begin{figure*}[hbt!]
  \includegraphics[width=0.97\linewidth]{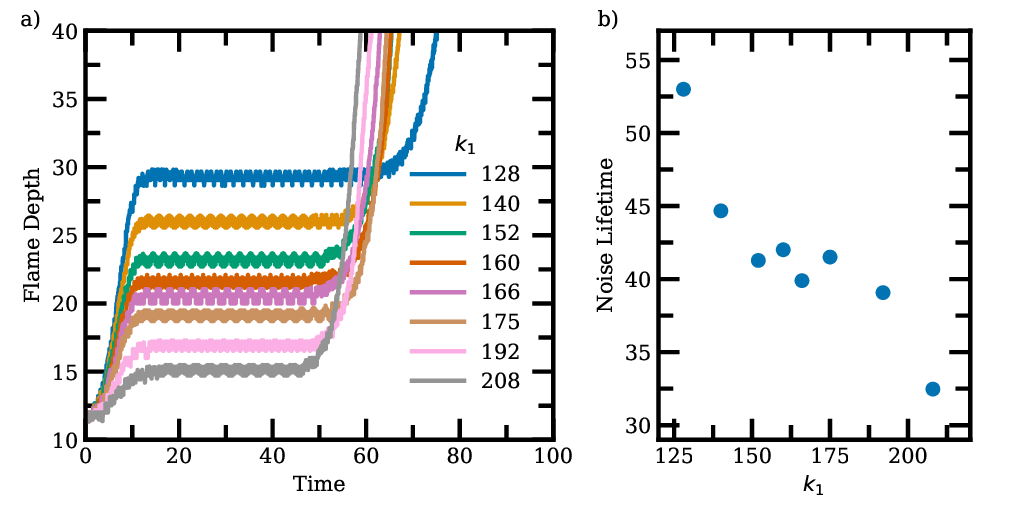}
\caption{(a) Single-Mode Simulations: the First Metastable Stage is eventually disrupted by numerical multimode noise. The flame depth is measured with the point method using $T=[0.005, 0.995]$ (see Appendix \ref{depth}). (b) Noise Lifetimes: the amount of time that the First Metastable Stage survives before it is disrupted by noise.}
\label{fig-single-mode}
\end{figure*}

\begin{figure*}[hbt!]
  \includegraphics[width=0.97\linewidth]{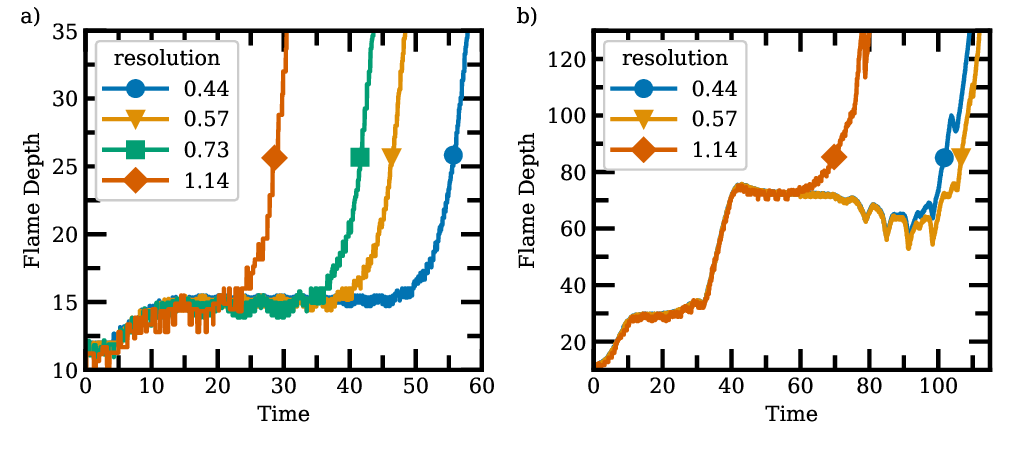}
\caption{(a) Single-Mode Resolution Study ($k_1=208$): the noise lifetime increases as the resolution is improved. (b) Two-Mode Resolution Study ($k_1=128$, $k_2=64$): the duration of the First Metastable Stage and the transition to and development of the Second Metastable Stage are well-converged. The Second Metastable Stage is broken by numerical multimode noise.}
\label{fig-resolution-checks}
\end{figure*}

First, we show that the lifetime of the First Metastable Stage increases as the resolution is improved. In an ideal single-mode simulation with no numerical error, we would expect the lifetime of the First Metastable Stage to be infinite. However, even in a well-resolved single-mode simulation, the First Metastable Stage is eventually destroyed by numerical multimode noise. Figure \ref{fig-single-mode}a shows the development of the First Metastable Stage and its destruction by noise for a variety of single-mode wavenumbers ($k_1$). Measured lifetimes are shown in Figure \ref{fig-single-mode}b. These ``noise lifetimes'' are the longest first metastable lifetimes that can be measured for each combination of resolution, $k_1$, and $A_1$. The noise lifetime for $k_1=128$ will be used in the next section to select an appropriate value for $A_2$. The noise lifetime should increase as the resolution is improved. Figure \ref{fig-resolution-checks}a demonstrates this good behavior using $k_1=208$ as an example: the First Metastable Stage plateau is longer when the resolution is better. Based on this resolution study, we decided to resolve the simulations as much as we could to allow the secondary perturbation more time to grow. Our chosen average resolution ($0.444$) is smaller than the measured viscous scale, implying that the simulations capture the energy cascade. The flame front itself should also be resolved because there are nearly 10 collocation (grid mesh) points across the distance ($4.394$) between the $T=0.1$ and $T=0.9$ contours of the laminar flame.

Second, we confirmed that the effect of the system perturbation disappears as the resolution is improved. When single-mode simulations are underresolved, the system perturbation mimics the effect of a secondary perturbation. Organized, large-scale structures emerge which are entirely due to numerical error. As the resolution improves, these structures disappear, showing that the amplitude of the system perturbation decreases. At our highest resolution, the system perturbation has little effect. However, we eliminate the system perturbation entirely in our two-mode simulations by choosing $k_1=128$, which aligns the bubbles of the flame front perfectly with the spectral elements.

\subsection{Two-Mode Simulations and Numerical Multimode Noise}

In this paper, we made the choice to show and discuss all of our two-mode simulations, even when they are affected by numerical multimode noise. Our hope is that this additional transparency will help future researchers in this area to plan and interpret their own simulations. In this section, we expand on the discussion in the main body of the paper and carefully consider the role of numerical multimode noise in our simulations.

First, we needed to ensure that the secondary perturbation actually played a role in breaking the First Metastable Stage and that we weren't just measuring the effects of numerical multimode noise. We did this by choosing $A_2$ appropriately. $A_2$ must be large enough that the lifetime of the First Metastable Stage of the simulation with the slowest growing secondary perturbation ($k_2=1$) is shorter than the noise lifetime of the primary perturbation, $k_1=128$. We chose $A_2=0.001$ because the lifetime of the First Metastable Stage of $k_2=1$ is slightly less than the noise lifetime (see Figure \ref{fig2}b). This choice allowed us to explore the full range of possible solutions from noise-dominated to secondary-perturbation-dominated by increasing $k_2$ (see Section \ref{latestage}).

Next, we identified the values of $k_2$ for which the First Metastable Stage is broken by true two-mode coupling between the primary and secondary. In Section \ref{latestage}, we show that the secondary outcompetes numerical multimode noise and breaks the First Metastable Stage on its own when $k_2 \ge 8$. This is visually apparent for simulations with GCD$(k_1=128, k_2) \ge 2$ which break up symmetrically.

Finally, we checked that the evolution of the $k_2 \ge 8$ simulations continued to be dominated by two-mode coupling until a GCD number of structures developed. Visually, the simulations must maintain symmetry until GCD identical structures develop. For example, Figure \ref{firstmetastablemerging} shows that the $k_2=210$ (GCD=$2$) simulation maintains symmetry until two giant structures form. 

After GCD identical structures form, symmetry will eventually be broken by the slowly growing numerical multimode noise. Figure \ref{firstmetastablepulsating} shows this process for $k_2=224$ (GCD=$32$). After 32 structures form (panel 4), the inverse k-cascade driven by two-mode coupling between the primary and secondary is complete. The symmetry breaking shown in panels 5-7 is due to numerical multimode noise and is not physical. Likewise, Figure \ref{secondmetastablemerging} for $k_2=64$ (GCD=$64$) shows the development of 64 identical bubbles which persist for some time before symmetry is broken by multimode noise in panels 8-9. Figure \ref{fig-resolution-checks}b shows a resolution study for this case. The duration of the First Metastable Stage and the transition to and development of the Second Metastable Stage are well-converged. The duration of the Second Metastable Stage does depend on resolution because the Second Metastable Stage is broken by numerical multimode noise instead of by a real physical perturbation. 

Ultimately, multimode noise affects simulations differently depending on their value of $k_2$. When $k_2 < 8$, multimode noise competes with the secondary perturbation and plays a role in breaking the First Metastable Stage. If these simulations could be carried out at higher resolution, we would expect the $k_2=1$ simulation to converge to the Nearly Symmetric Merging solution type and simulations with $k_2=2,4,6$ to converge to the First Metastable Merging solution type.  These simulations are unphysical in the sense that they show behavior that is not due to two-mode coupling, but their sort of asymmetrical behavior is likely in the real world where ``noisy'' multimode perturbations are ubiquitous. For simulations with $k_2 \ge 8$, the solution type is converged. Simulations with GCD$\ge2$ remain symmetric until a GCD number of identical structures develop. Eventually, multimode noise breaks this symmetry. Increasing the resolution of these simulations would likely delay the symmetry breaking, but would not change the solution type. Simulations with GCD=$1$ always appear asymmetrical, but they are still converged (to the Nearly Symmetric Merging solution) as long as $k_2 \ge 8$. Overall, solution types for $k_2 \ge 8$ are converged and our conclusions about two-mode coupling are robust.

\vspace{-0.25cm}
\section{Flame Depth and Bubble Depth Measurements} \label{depth}

During the development of the Rayleigh-Taylor instability, distinctive structures known as ``bubbles'' and ``spikes'' form on the flame front. Bubbles are lighter ashes moving upwards, while spikes are heavier fuels moving downwards. To quantify the growth of these structures, we measure the positions of the top of the bubbles and the bottom of the spikes using two methods: the point method and the profile method. Both methods use temperature thresholds, for example $T=(0.1,0.9)$, to define the top and bottom of the flame. The flame top corresponds to the highest point with temperature above the lower threshold, whereas the flame bottom corresponds to the lowest point with temperature below the upper threshold. The point method uses the vertical position of the single point that meets these criteria; the profile method first calculates the horizontal average of the temperature field and then identifies the vertical position that satisfies the thresholds. 

\begin{figure*}[hbt!]
\centering
\includegraphics[width=0.98\textwidth]{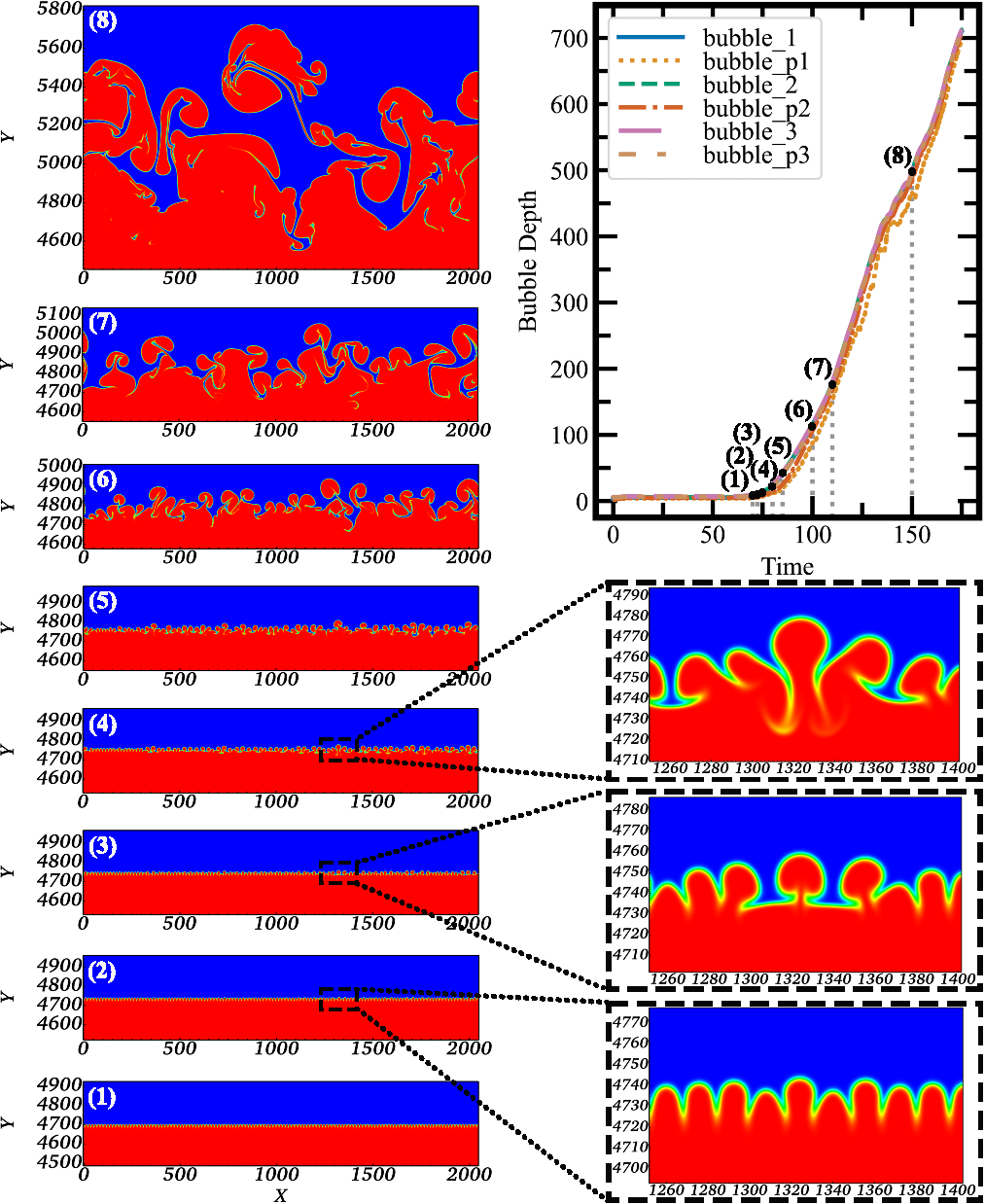}
\caption{Chaotic Burning Solution ($k_1=128$, $k_2=1$, GCD=$1$). See Supplemental Material Movie for run 316 \citep{liu-supplement}. \label{chaoticburning}}
\end{figure*}

\begin{figure*}[hbt!]
\centering
\includegraphics[width=1\textwidth]{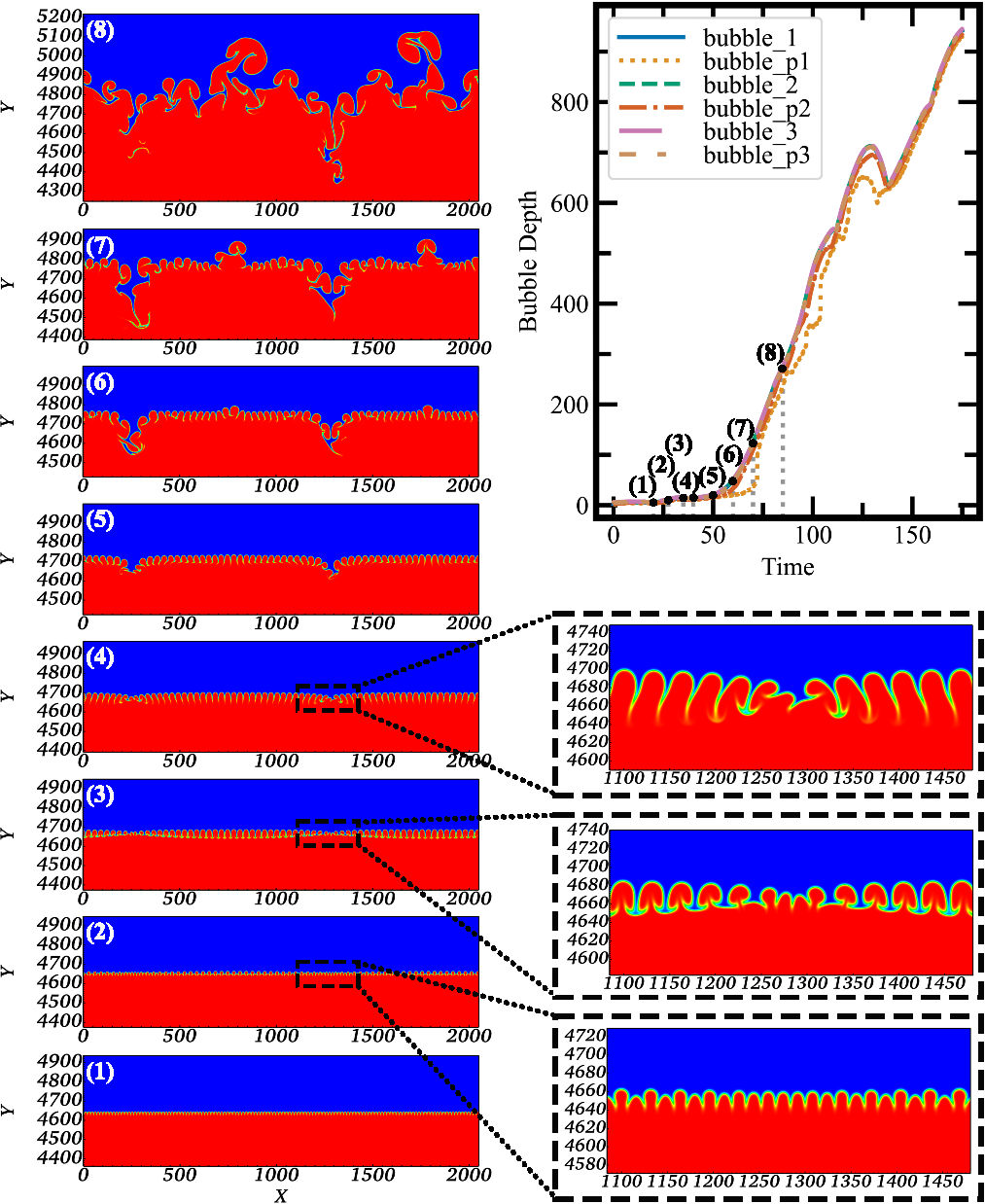}
\caption{Nearly Symmetric Merging Solution ($k_1=128$, $k_2=65$, GCD=$1$).  See Supplemental Material Movie for run 322 \citep{liu-supplement}.}
\label{nearlysymmetricmerging}
\end{figure*}

\begin{figure*}[hbt!]
\centering
\includegraphics[width=1\textwidth]{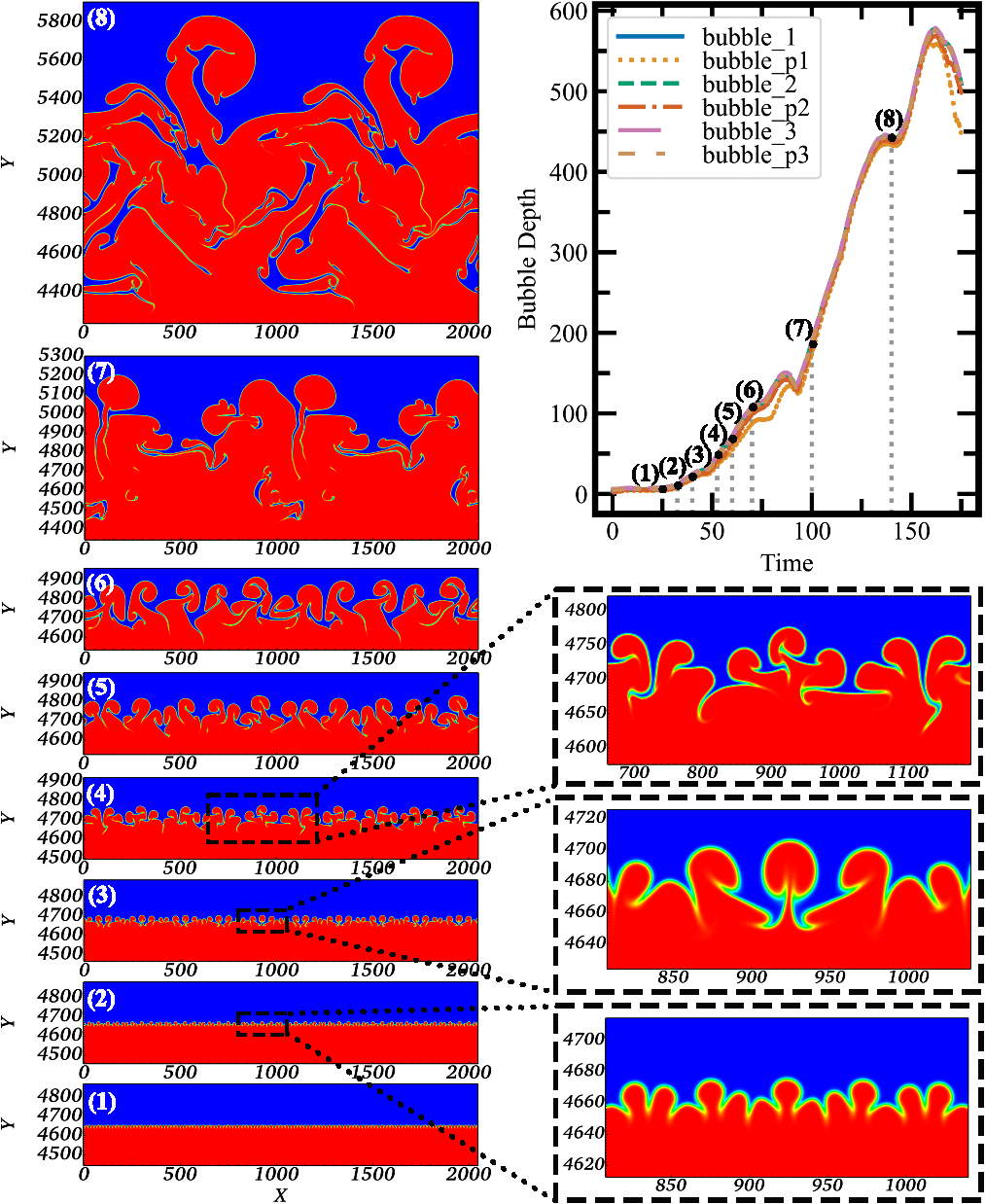}
\caption{First Metastable Merging Solution ($k_1=128$, $k_2=210$, GCD=$2$). See Supplemental Material Movie for run 337 \citep{liu-supplement}.}
\label{firstmetastablemerging}
\end{figure*}

\begin{figure*}[hbt!]
\centering
\includegraphics[width=1\textwidth]{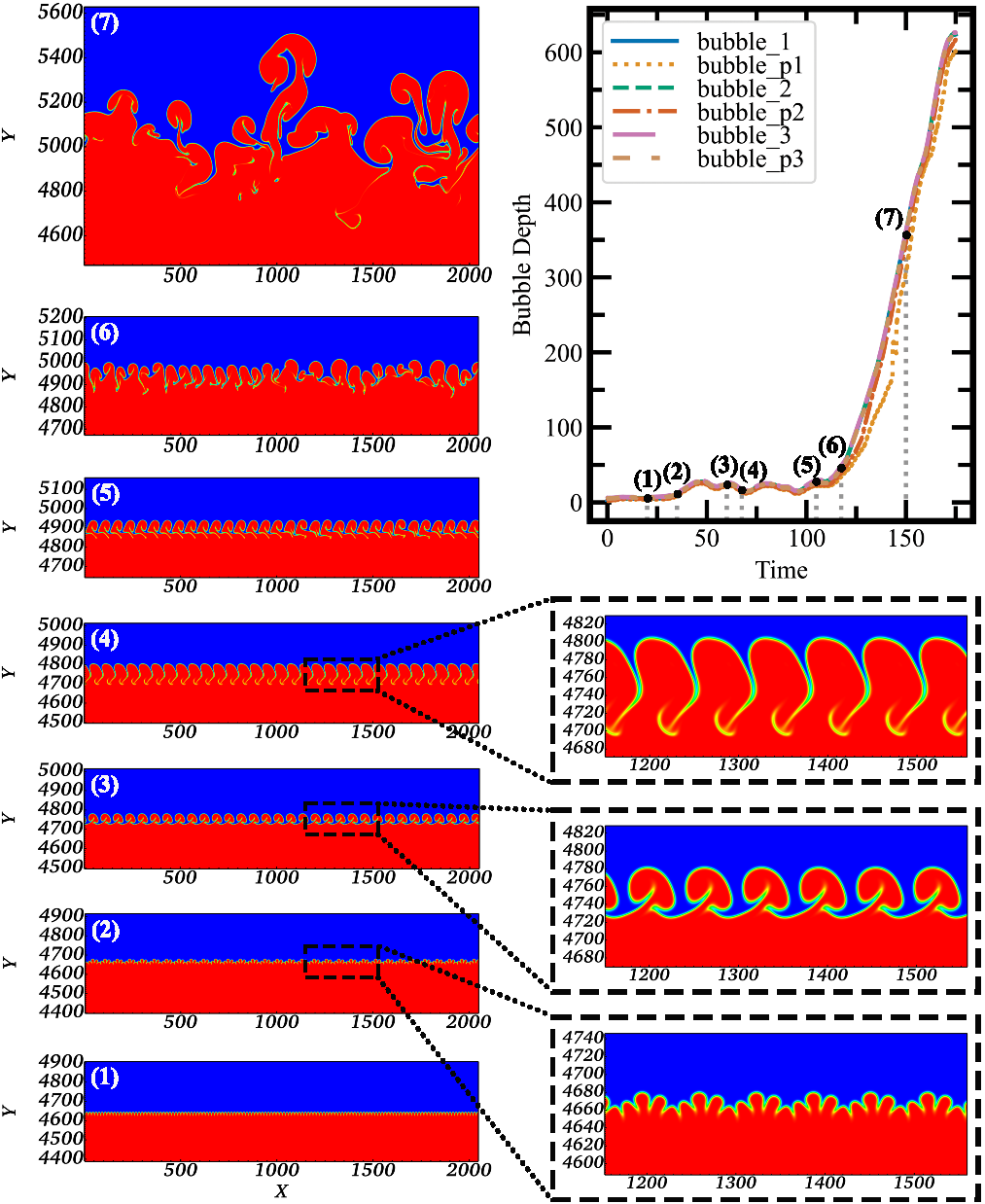}
\caption{First Metastable Pulsating Solution ($k_1=128$, $k_2=224$, GCD=$32$). See Supplemental Material Movie for run 347 \citep{liu-supplement}.}
\label{firstmetastablepulsating}
\end{figure*}

\begin{figure*}[hbt!]
\centering
\includegraphics[width=1\textwidth]{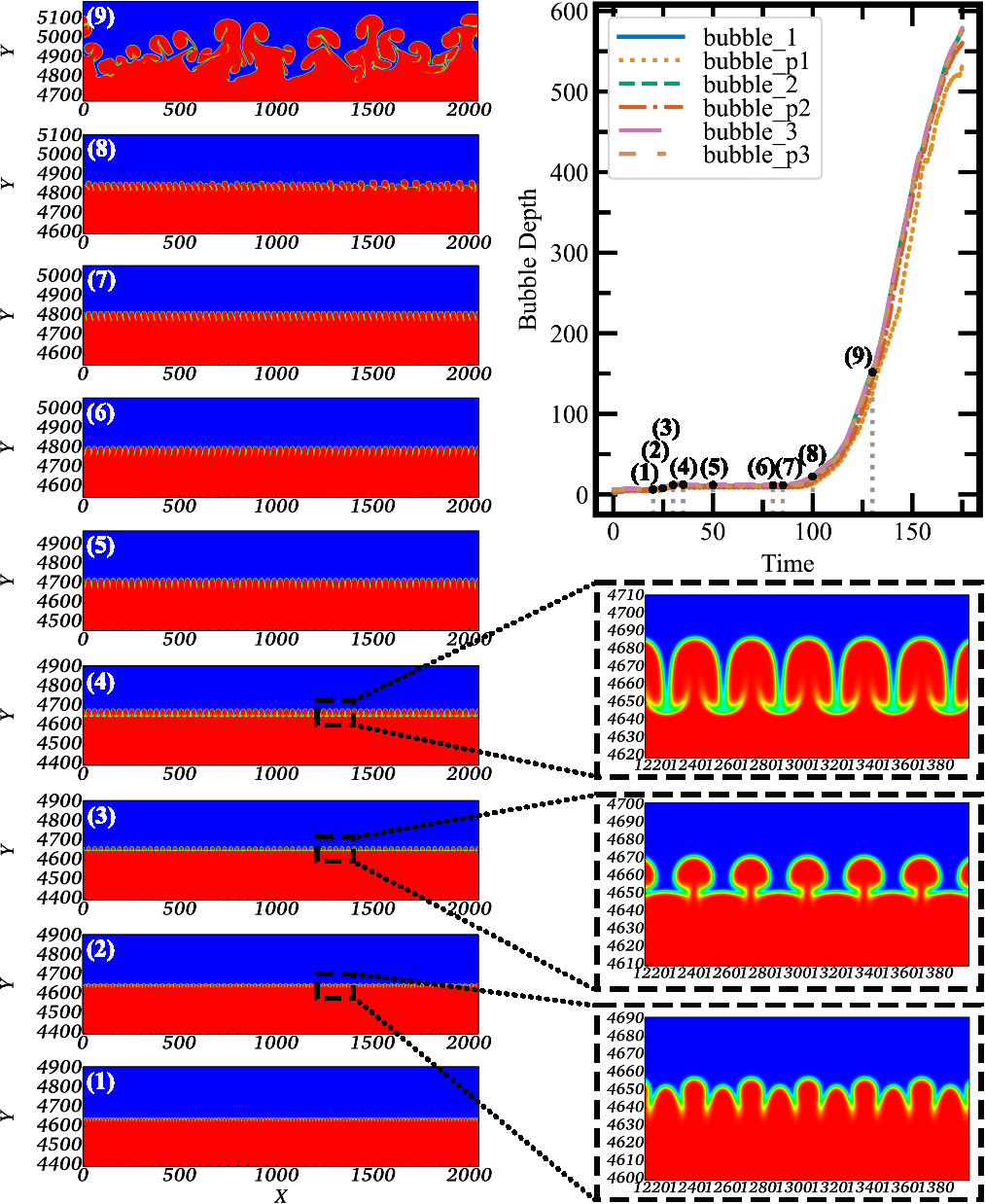}
\caption{Second Metastable Merging Solution ($k_1=128$, $k_2=64$, GCD=$64$). See Supplemental Material Movie for run 315 \citep{liu-supplement}.}
\label{secondmetastablemerging}
\end{figure*}

To turn these measurements into mixing layer height measurements, we take two approaches. First, we compute the size of the entire mixing layer by subtracting the spike position from the bubble position. This measurement includes both bubbles and spikes and we call it the ``flame depth''. We also measure a ``bubble depth'' that excludes the spikes by subtracting the average position for the flame (calculated by integrating the flame speed over time) from the bubble position. The insets in Figures \ref{chaoticburning}-\ref{secondmetastablemerging} show measurements of the bubble depth made using both the point and profile methods across the three threshold sets: $T=(0.1,0.9)$, $(0.05,0.95)$, and $(0.005,0.995)$, denoted respectively by numbers 1, 2, and 3 in the plot legends. The profile method measurements are labeled by the letter `p' in front of the number. 
\clearpage

\section{Late Stage Solution Visualizations} \label{solutions}
\vspace{-0.3cm}
\clearpage

\clearpage

\clearpage

\clearpage


%

\end{document}